\documentclass[apj]{emulateapj}

\shorttitle{Coma Cluster Outskirts}
\shortauthors{Simionescu et al.}

\begin{document}

\title{Thermodynamics of the Coma Cluster Outskirts}

\author{A. Simionescu\altaffilmark{1,2,3}, N. Werner\altaffilmark{1,2}, O. Urban\altaffilmark{1,2,4}, S. W. Allen\altaffilmark{1,2,4}, A. C. Fabian\altaffilmark{5}, A. Mantz\altaffilmark{6,7}, K. Matsushita\altaffilmark{8},\\ P. E. J. Nulsen\altaffilmark{9}, J. S. Sanders\altaffilmark{5,10}, T. Sasaki\altaffilmark{8}, T. Sato\altaffilmark{8}, Y. Takei\altaffilmark{3}, S. A. Walker\altaffilmark{5}}
\affil{$^1$KIPAC, Stanford University, 452 Lomita Mall, Stanford, CA 94305, USA}
\affil{$^2$Department of Physics, Stanford University, 382 Via Pueblo Mall, Stanford, CA  94305-4060, USA}
\affil{$^3$Institute of Space and Astronautical Science (ISAS), JAXA, 3-1-1 Yoshinodai, Chuo-ku, Sagamihara, Kanagawa, 252-5210 Japan}
\affil{$^4$SLAC National Accelerator Laboratory, 2575 Sand Hill Road, Menlo Park, CA 94025, USA}
\affil{$^5$Institute of Astronomy, Madingley Road, Cambridge CB3 0HA, UK}
\affil{$^6$Kavli Institute for Cosmological Physics, University of Chicago, 5640 South Ellis Avenue, Chicago, IL 60637, USA}
\affil{$^7$Department of Astronomy and Astrophysics,University of Chicago, 5640 South Ellis Avenue,Chicago, IL 60637-1433, USA}
\affil{$^8$Department of Physics, Tokyo University of Science, 1-3 Kagurazaka, Shinjuku-ku, Tokyo 162-8601, Japan}
\affil{$^9$Harvard-Smithsonian Center for Astrophysics, 60 Garden St., Cambridge, MA 02138, USA}
\affil{$^{10}$Max Planck Institute for Extraterrestrial Physics, Giessenbachstr 1, 85748 Garching, Germany}

\begin{abstract}

We present results from a large mosaic of Suzaku observations of the Coma Cluster, the nearest and X-ray brightest hot ($\sim$8~keV), dynamically active, non-cool core system, focusing on the thermodynamic properties of the ICM on large scales. 
For azimuths not aligned with an infalling subcluster
towards the southwest, our measured temperature and X-ray brightness 
profiles exhibit broadly consistent radial trends, with the temperature decreasing from about 8.5~keV at the cluster center to about 2~keV at a radius of 2~Mpc, which is the edge of our detection limit. The SW merger significantly boosts the surface brightness, allowing us to detect X-ray emission out to $\sim$2.2 Mpc along this direction.
Apart from the southwestern infalling subcluster, the surface brightness profiles show multiple edges around radii of 30--40 arcmin. The azimuthally averaged temperature profile, as well as the deprojected density and pressure profiles, all show a sharp drop consistent with an outward propagating shock front located at 40 arcmin, corresponding to the outermost edge of the giant radio halo observed at 352 MHz with the WSRT. The shock front may be powering this radio emission. 
A clear entropy excess inside of $r_{500}$ reflects the violent merging events linked with these morphological features.
Beyond $r_{500}$, the entropy profiles of the Coma Cluster along the relatively relaxed directions are consistent with the power-law behavior expected from simple models of gravitational large-scale structure formation. The pressure is also in agreement at these radii with the expected values measured from SZ data from the Planck satellite. 
However, due to the large uncertainties associated with the Coma Cluster measurements, we cannot yet exclude an entropy flattening in this system consistent with that seen in more relaxed cool core clusters.

\end{abstract}


\section{Introduction}\label{intro}

All of the matter in the Universe, both luminous and dark, is distributed in a complex structure of sheets, filaments and voids which evolved following the gravitational collapse of small perturbations in the primordial density field. Clusters of galaxies are located at the nodes of this filamentary structure, also known as the cosmic web, and are constantly evolving and growing by accreting matter from the surrounding large-scale structure. 

The outskirts of galaxy clusters present an opportunity to study this process as it happens. The virial radius\footnote{We adopt as our definition of the virial radius r$_{200}$, within which the mean total density is 200 times the critical density of the universe at the redshift of the cluster} approximately marks the border between regions of equilibration and in-fall. Merger activity from accreting material plays an important role in the dynamics of the outer regions. In order to understand large-scale structure formation in detail, it is therefore necessary to study the thermodynamic properties of the intra-cluster medium close to the virial radius in systems with a wide range of ages/masses and dynamical states.

Surface brightness profiles out to the virial radius have been studied using ROSAT data \citep{Vikhlinin99,Neumann05}, which provided density estimates but no information about the ICM temperature. Using archival data from ROSAT, \citet{eckert2012} find that galaxy clusters deviate significantly from spherical symmetry beyond $\sim r_{500}$, with only small differences between relaxed and disturbed systems.

Robust measurements of the spectroscopic properties of the faint cluster outskirts have been made possible only recently, owing to the low and stable background of the Japanese-US Suzaku satellite. However, results to date have been obtained primarily for massive, relaxed clusters, such as PKS 0745-191 \citep{George09,Walker0745}, Abell 2204 \citep{Reiprich09}, Abell 1795 \citep{Bautz09}, Abell 1413 \citep{Hoshino10}, the Perseus Cluster \citep{SimionescuSci}, and Abell 2029 \citep{Walker2029}. By comparison, fewer results are available for less massive and/or less dynamically relaxed systems. For example, Abell 2142 \citep{Akamatsu11} and Abell 3667 \citep{Akamatsu12} are so far the only clear mergers for which the thermodynamic profiles have been investigated out to $r_{200}$. Measurements of the properties of cluster outskirts have also been published for Abell 1689 \citep{Kawaharada10}, thought to be a line-of-sight merger; the Virgo Cluster \citep{Urban11}, a young, dynamically unrelaxed, low-mass system; the Hydra~A \citep{Sato12} and Centaurus clusters \citep{Walker13}, which are low-mass, relatively relaxed systems; and for the relaxed fossil group/poor cluster RXJ1159+5531 \citep{Humphrey12}. We refer the reader to \citet{Reiprich13} for a more comprehensive review.

Here, we present results from a Suzaku Large Project targeted on the Coma Cluster, which is the nearest and X-ray brightest hot ($\sim$8~keV), dynamically active, non-cool core system \citep[e.g.][]{edge1990,white1993,briel2001,neumann2003}.

We assume a $\Lambda$CDM cosmology with $\Omega_{\rm m}$=0.27, $\Omega_\Lambda$=0.73, and H$_0$=70 km/s/Mpc. At the redshift of the cluster, z=0.0231 \citep{Struble99}, one arcminute corresponds to 28 kpc.

\section{Observations}

\subsection{Suzaku}

\begin{figure*}
\begin{center}
\includegraphics[width=0.97\columnwidth]{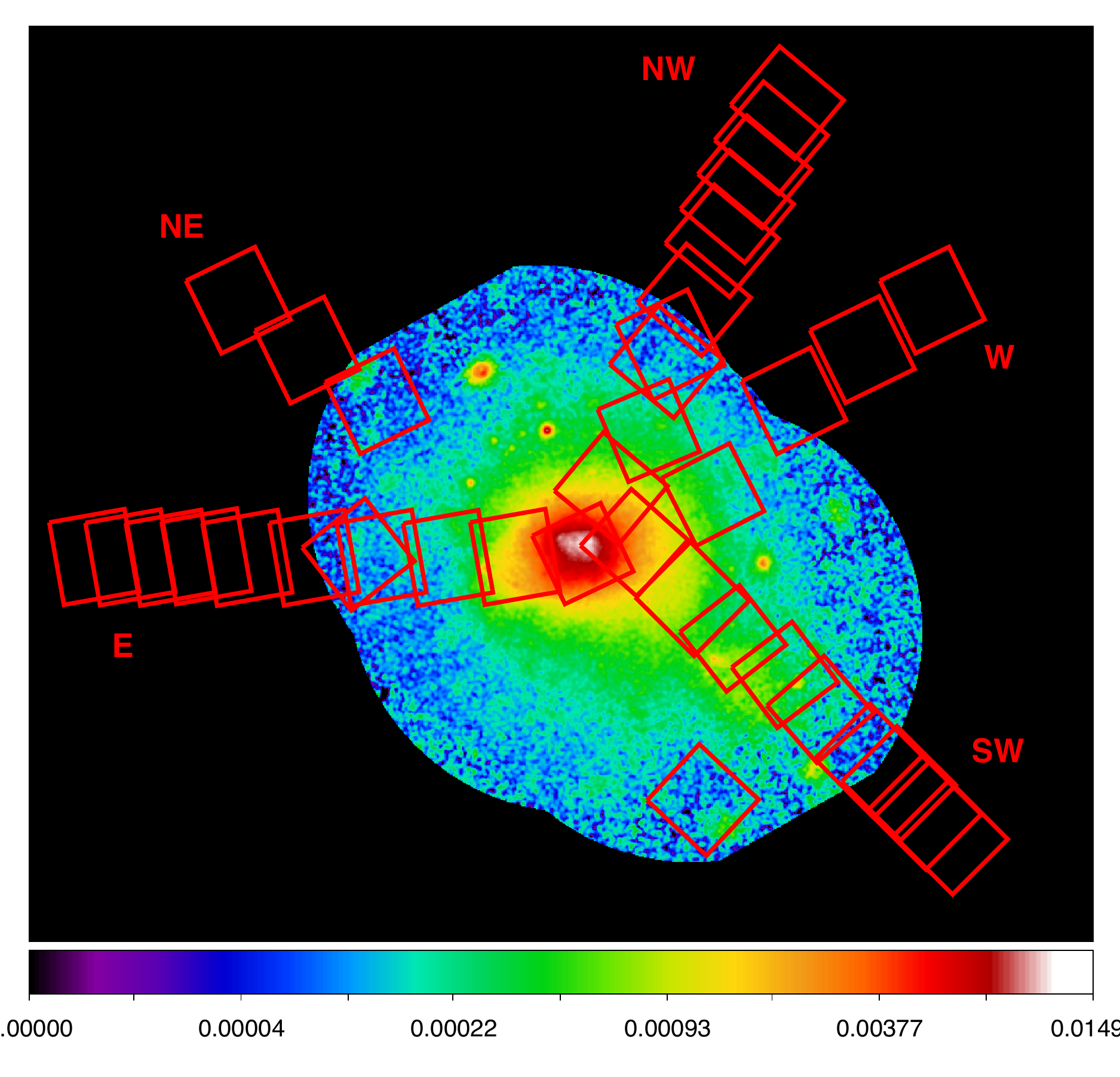}
\includegraphics[width=0.95\columnwidth]{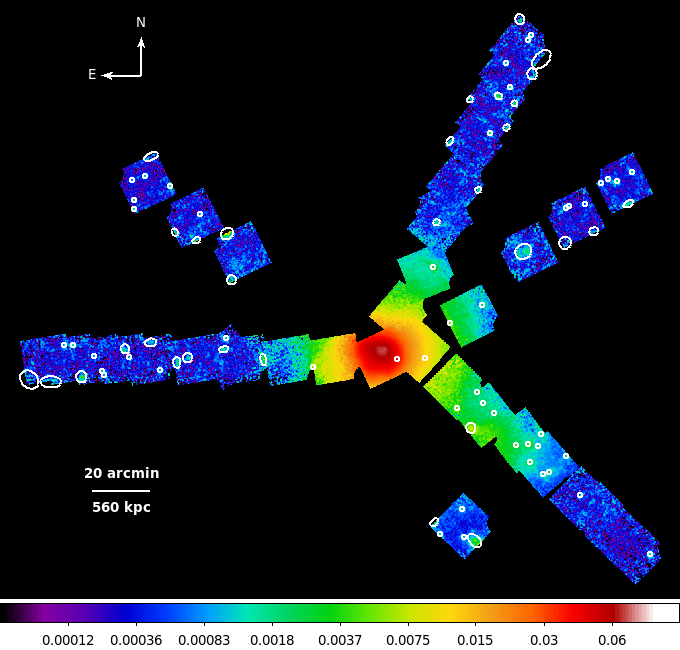}
\end{center}
\caption{{\it Left:} ROSAT-PSPC image of the Coma Cluster in the 0.7--2 keV energy band; the locations of the Suzaku mosaic pointings are shown as red squares. {\it Right:} X-ray surface brightness map of the Coma Cluster from the Suzaku mosaic in the 0.7--7 keV energy band. Point sources and other artifacts and structures excluded from the spectral analysis are shown in white. \label{coma_image}}
\end{figure*}

A large mosaic of observations of the Coma Cluster was obtained as a Suzaku Large Project during AO-6, covering the E, NW and SW directions contiguously out to a radius of $2^\circ$ (PI A. Simionescu). The mosaic consists of 24 pointings with clean exposure times between 5 and 20~ks each, for a total observing time of 295~ks.  
In addition, 6 more pointings were awarded during the same AO as a Guest Observer program (PI T. Sato), covering regions located towards the NE and W from the cluster center, with a total effective exposure of 175~ks. We combined these pointings with archival data obtained in AOs 1-5, for a total of 1.05~Ms of observing time. The details of all the pointings used in this work are summarized in Table \ref{tab:observations}. We label the pointings with letters indicating the direction of the arm that the pointing belongs to, followed by numerical indices which increase with increasing radius, such that e.g. all pointings with index 4 are offset by approximately the same distance from the cluster core. Pointings labeled with half-integer indices are marked such that for example 3.5 lies between and partly overlaps with pointings 3 and 4.
The spatial coverage of our combined data set is shown in Figure \ref{coma_image}.   

The data were reduced using the tools available in the HEAsoft package (version 6.12) to create a set of cleaned event lists with hot or flickering pixels removed. All standard recommended screening criteria were applied \footnote{Arida, M., XIS Data Analysis, http://heasarc.gsfc.nasa.gov/docs/suzaku/analysis/abc/node9.html (2010)}. We only included observation periods with the geomagnetic cut-off rigidity COR $>$ 6 GV. The X-ray surface brightness image, extracted in the 0.7--7 keV energy band and corrected for vignetting and instrumental background, is shown in the right panel of Figure \ref{coma_image}. 

We used the latest calibration files which account for the modified non X-ray background of the XIS1 detector following the increase in the charge injection level of 2011 June 1; in addition, for the XIS1 spectral analysis, we excluded two columns adjacent on each side to the charge-injected columns (the standard is to exclude one column on either side). This is because the injected charge may leak into these additional columns and cause an increase in the instrumental background. We applied the latest contamination layer calibration from 2012 July 19.
 
\subsection{ROSAT}

The Coma Cluster was also covered by a mosaic of four ROSAT Position Sensitive Proportional Counter (PSPC) observations extending out to radii of approximately $60^\prime$ (with observation identifiers rp800006N00, rp800013N00, rp800005N00, and rp800009N00). These observations were performed between 1991 June 16--18, and their combined clean exposure time is 78~ks.
The data were reduced using the Extended Source Analysis Software (ESAS, \citealt{Snowden1994}). Background and exposure corrected images in three energy bands (0.7-0.9, 0.9-1.3, 1.3-2.0 keV respectively) were combined, having removed artifacts associated with the detector edges. The resulting image is shown in the left panel of Figure \ref{coma_image}.

\section{Spectral modeling}

We used the Suzaku mosaic to extract spectra from annuli centered on the cluster center, $(\alpha,\delta)$=(12:59:42.44,+27:56:45.53). The projected and deprojected profiles of thermodynamic properties were obtained with the XSPEC \citep[version 12.7,][]{arnaud1996} spectral fitting package, employing the modified C-statistic estimator. We used the {\small projct} model to deproject the data under the assumption of spherical symmetry. Sets of spectra extracted from concentric annuli in the 0.7--7.0 keV band were modeled simultaneously. We modeled each shell as a single-temperature thermal plasma in collisional ionization equilibrium using the {\it apec} code \citep{smith2001}, with the temperature and spectrum normalization as free parameters. Unless otherwise noted, the metal abundance was set to 0.3 of the Solar value in the units of \cite{grevesse1998}. The Galactic absorption column density was fixed to the average value measured at the location of each respective Suzaku field from the Leiden-Argentine-Bonn radio HI survey \citep{kalberla2005}; the values used are given in Table \ref{tab:observations}.

\subsection{Background subtraction}\label{cxb_default}

The cosmic X-ray background model was obtained by fitting spectra from regions free of cluster emission located beyond 80 arcmin from the cluster core, as well as two ROSAT All-Sky Survey (RASS) spectra extracted from $1^\circ$-radius circles centered $3^\circ$ from the cluster core towards the E and NW. Due to the presence of a galaxy group (SDSSCGB 51748) as well as several background (z$\sim$0.2) clusters, including Abell 1632, at radii of around 90--100 arcmin towards the W, we have not included the western pointings in our main background analysis.  

The background model consisted of a power-law emission component that accounts for the unresolved population of point sources, one absorbed thermal plasma model for the Galactic Halo (GH) emission, and an unabsorbed thermal plasma model for the Local Hot Bubble (LHB). We first fitted the Suzaku spectra in the hard band (2--7 keV) and determined the power-law index, for which we obtain $1.50\pm0.04$. This is slightly steeper than the widely used value of 1.41 reported previously by e.g. \cite{deLuca04} from XMM-Newton MOS data. 

Next, we fitted the Suzaku and RASS spectra in parallel, but using chi-square minimization for the RASS and C-statistic for the Suzaku spectra. We used the 0.1--2 keV band for the RASS spectra and 0.7--7 keV band for the Suzaku spectra. For the RASS data, we fixed the power-law parameters to the values reported by \cite{kuntz2000} ($\Gamma=1.46$, $Y=8.88\times10^{-7}$ photons keV$^{-1}$ cm$^{-2}$ s$^{-1}$ arcmin$^{-2}$ at 1 keV) while for the Suzaku spectra, we fixed the power-law index to the value measured in the Suzaku hard band ($\Gamma=1.50$) and left the normalization free. The temperatures and normalizations of the GH and LHB components were left free in the fit, with the metallicities fixed to 0.3 and 1 solar, respectively. The results are summarized in Table \ref{cxb}. 

\begin{table}
\caption{Cosmic X-ray Background parameters. Errors are given at the $\Delta$C=1 level.}
\begin{center}
\begin{tabular}{lcc}
\hline
 	& $kT$ (keV) or $\Gamma$ & $Y^\dagger$  \\
\hline
CXB & 1.50 & $1.15\pm0.01$ \\
GH & $0.25\pm0.01$ & $2.9\pm0.3$\\
LHB & $0.10\pm0.002$ & $1.6\pm0.05$ \\
\hline
\end{tabular}
\end{center}
$^\dagger$spectrum normalization, $\int n_e n_H dV$, in units of $\frac{10^{-17}}{4\pi [D_A (1+z)]^2}$ cm$^{-5}$ per $20^2\pi$ sq arcmin area. 
\label{cxb}

\end{table} 

\begin{figure}
\begin{center}
\includegraphics[width=1.1\columnwidth,bb=18 18 592 490]{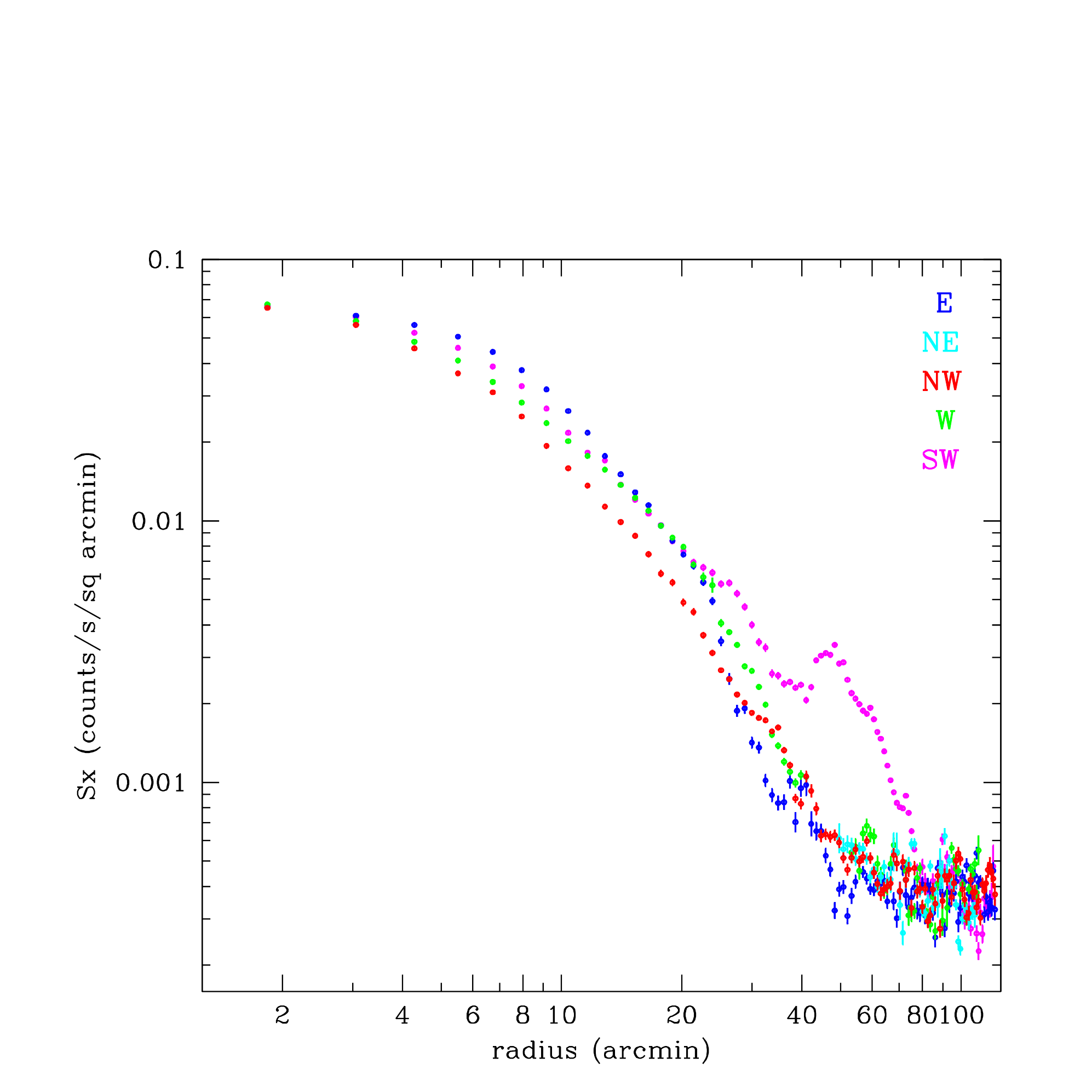}
\end{center}
\caption{Surface brightness profiles in the 0.7--7 keV band obtained with Suzaku. \label{sbprof}}
\end{figure}

\section{Results}

\subsection{Surface brightness features}

It is well known that the Coma Cluster is undergoing a merger, with a bright subcluster located towards the south-west of the main cluster center readily apparent in the ROSAT and Suzaku images. This ongoing merger clearly affects the surface brightness distribution, which is significantly higher at radii between 25--80 arcmin along the merger axis compared to the other more relaxed directions observed with Suzaku (see Fig.~\ref{sbprof}). 

Further deviations from spherical symmetry are observed among the other, relatively less disturbed directions. 
Both the E and W directions show an excess in X-ray emission compared to the NW arm at small radii (4--30 arcmin). Along the E arm, we observe a strong surface brightness gradient between 25--35 arcmin; an additional surface brightness bump followed by a drop is seen at $\sim$40~arcmin along this direction. Along the W arm, the surface brightness also shows a steep gradient around 25--35 arcmin, though less steep than for the E arm; the NW profile shows a marked boost between 30--40 arcmin, bringing it in agreement with the W brightness at these radii. 

It is important to note that, beyond a radius of approximately 80 arcmin, the surface brightness profiles along all the 5 different directions probed by our Suzaku observations converge to very similar levels. This indicates that the surface brightness of the cosmic X-ray background is roughly constant across the field of the mosaic.

\subsection{Thermodynamic properties of the ICM}

\subsubsection{Relatively Undisturbed Directions}

\begin{figure}
\begin{center}
\includegraphics[width=\columnwidth]{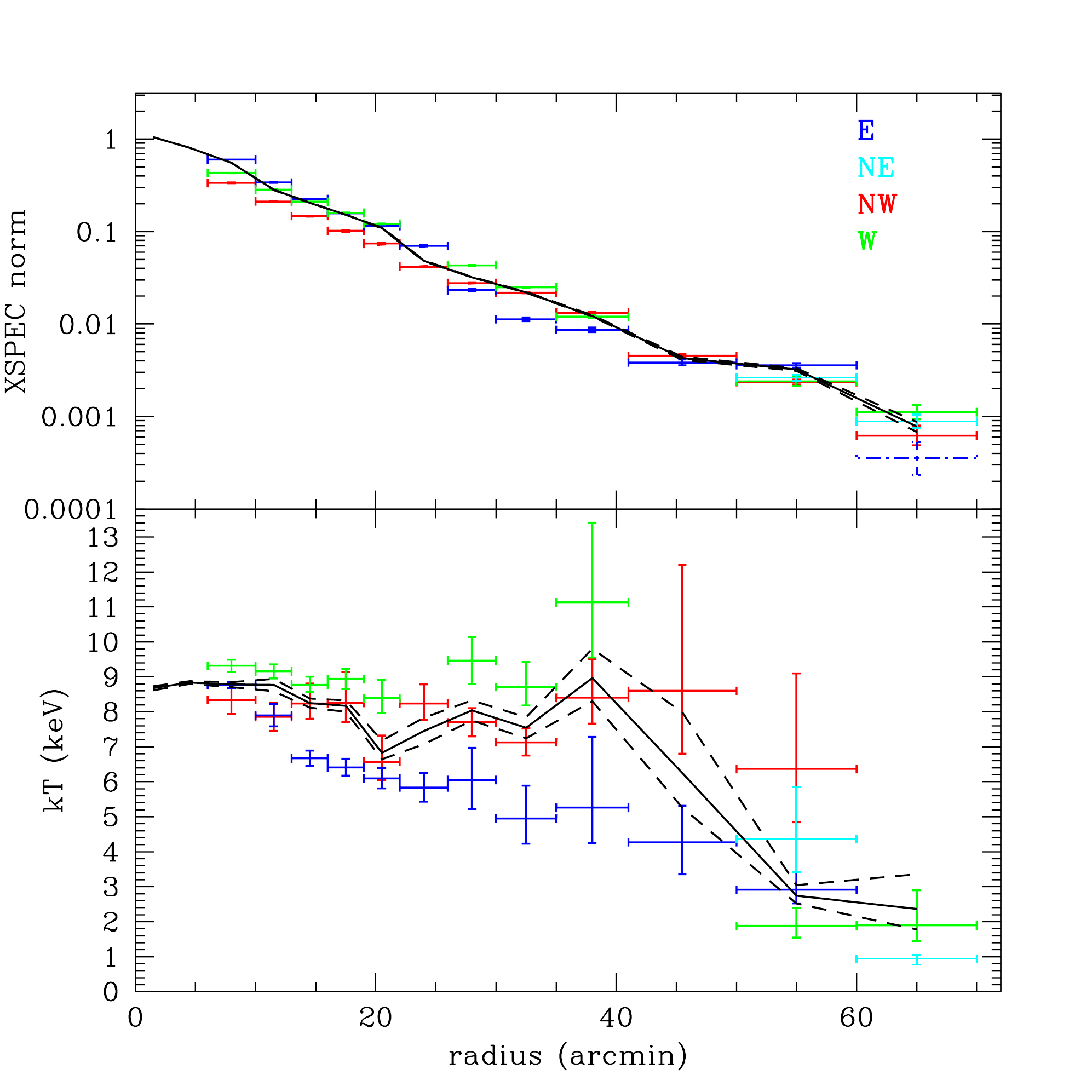}
\end{center}
\caption{Projected radial profiles of the temperature and spectrum normalization obtained from the Suzaku data along azimuths not aligned with the infalling southwestern subcluster. The average profiles and their 1$\sigma$ confidence intervals are shown with solid and dashed black lines, respectively. The spectrum normalization, $\int n_e n_H dV$, is given in units of $\frac{10^{-14}}{4\pi [D_A (1+z)]^2}$ cm$^{-5}$ per $20^2\pi$ sq arcmin area. \label{projrel}}
\end{figure}

In Figure \ref{projrel}, we show the radial profiles of projected temperature and spectral normalization obtained from our spectral analysis for the relatively undisturbed directions covered by the Suzaku mosaic, namely the E, NE, NW, and W. We detect the Coma Cluster along these directions out to a radius of up to 70 arcmin (2~Mpc). In the last annulus, spanning 60--70 arcmin, the temperature is constrained only along the W and NE directions; towards the NW and E, we have determined the best-fit spectral normalizations by fitting data from all 4 azimuths with a common average temperature. Towards the E, the spectral normalization in this last annulus is not robust to systematic uncertainties related to CXB fluctuations, as we will show in the next section, therefore this data point is shown with a dot-dashed line.

Both the temperature and spectral normalization are in broad agreement across the arms and show consistent radial trends among the different directions probed, with the temperature decreasing from about 8.5~keV at the cluster center to about 2~keV in the outermost annulus. An exception is the temperature along the eastern arm, which is systematically lower than towards the NW and W from the center of the cluster out to 50 arcmin (1.4 Mpc). In the same figure, we also show the azimuthally averaged profiles obtained by jointly fitting the spectra corresponding to the relatively undisturbed azimuths for a given radial bin with a common temperature and spectrum normalization. 

\begin{figure}
\begin{center}
\includegraphics[width=\columnwidth]{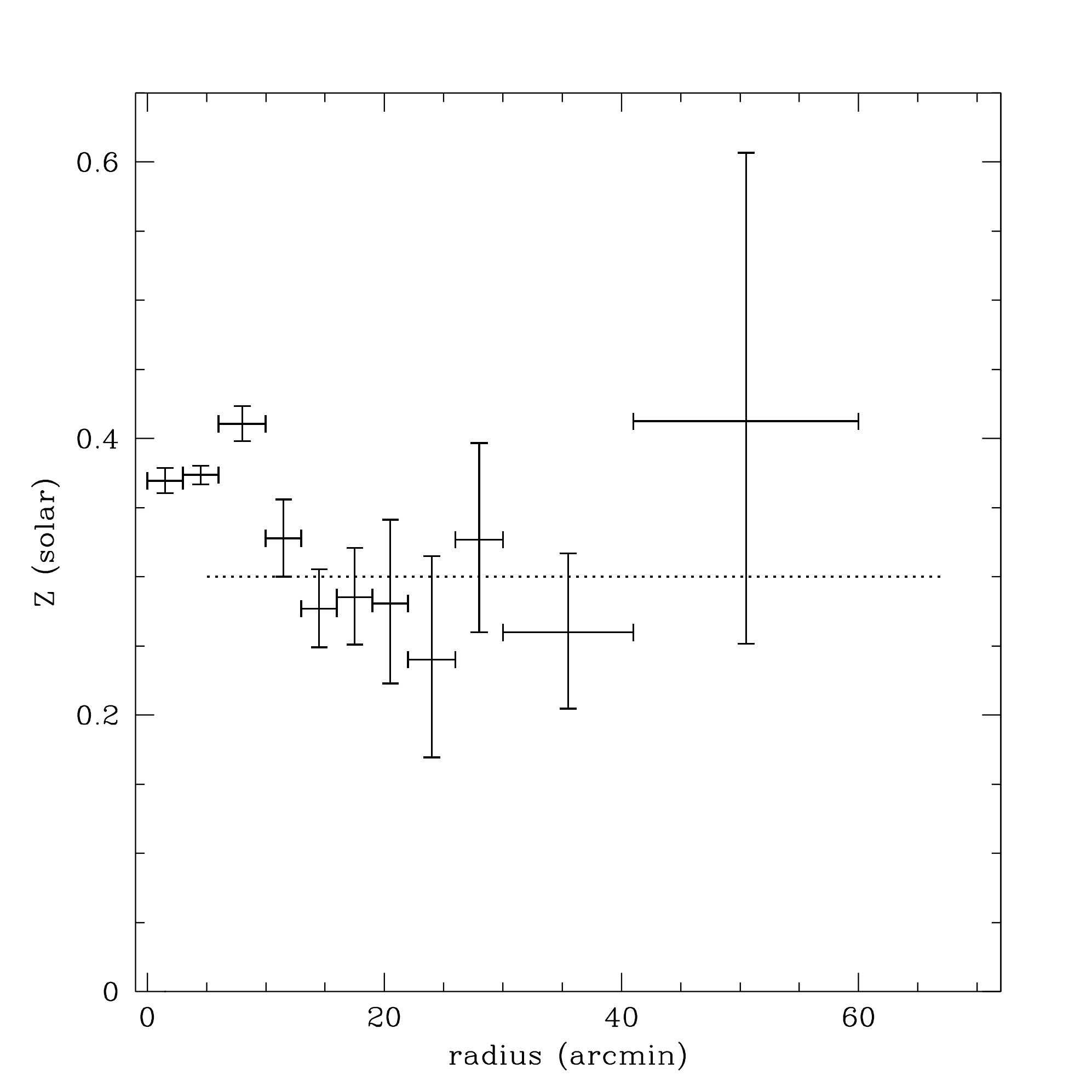}
\end{center}
\caption{Projected radial profile of the metallicity obtained from the Suzaku data along azimuths not aligned with the infalling southwestern subcluster. The horizontal line marks the value 0.3 solar. \label{metals}}
\end{figure}

We furthermore performed a fit to the spectra in each annulus where the metallicity was allowed to vary as a free parameter. To improve the statistical precision, we tied all the metallicities corresponding to a given radial bin to the same value (the temperatures and spectral normalizations were still allowed to vary with azimuth for a given annulus). The result is shown in Figure \ref{metals}. Apart from the last annulus (60--70 arcmin) and the cluster core, the best-fit values agree well with the value of 0.3 Solar which was used above. In the last annulus, by contrast, the $2\sigma$ upper limit on the metallicity is only 0.13 of the Solar value. This apparent drop in metallicity may result from multi-temperature structure in the ICM, which can cause the abundance measurement to be biased low especially at lower ICM temperatures, similar to what was found at large radii in the Virgo Cluster \citep{Urban11}.

The fact that the observed metallicity profiles are flat out to large radii supports the conclusion that the dominant enrichment mechanism are galactic winds, which were particularly strong and dominated the enrichment of the intergalactic medium at $z=2-3$. The ICM observed at large radii was thus most likely pre-enriched before it fell into the gravitational potentials of clusters, which explains the lack of a metallicity gradient as a function of radius (see Werner et al. 2013, submitted). 

\subsubsection{The South-Western Merger}

\begin{figure}
\begin{center}
\includegraphics[width=\columnwidth]{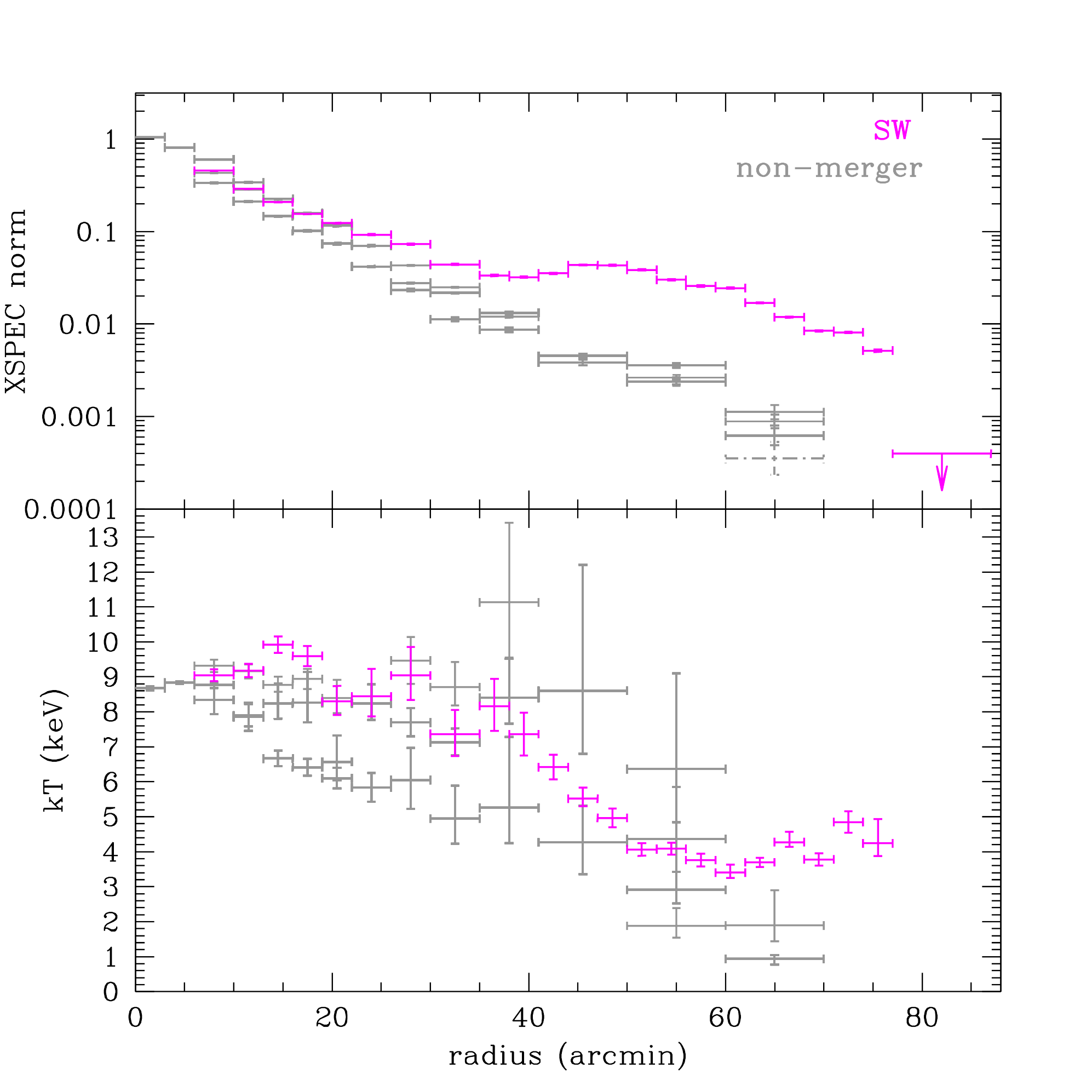}
\end{center}
\caption{Projected radial profiles of the temperature and spectrum normalization obtained from the Suzaku data along the southwestern direction, compared to other azimuths not aligned with the infalling subcluster. \label{projsw}}
\end{figure}

In Figure \ref{projsw}, we show the projected temperature and spectral normalization profiles along the direction of the infalling southwestern subcluster. The ongoing merger boosts the surface brightness distribution along this direction from about 25 arcmin (700 kpc) to the edge of the detection limit, with the peak excess located about 1.3~Mpc from the main cluster core. As a result of this boost, we detect X-ray emission out to larger radii than towards the other directions. The statistical error bars are also significantly shrunk because a very deep archival observation with an exposure time of 176~ks covers the radial range of about 63--78 arcmin.

The projected temperature profile shows a steady decline from about 35--55 arcmin, in agreement with the temperature decrease seen along the more relaxed W and NW directions. Beyond this, the ICM temperature remains practically constant over one Mpc in radius (from about 55--78 arcmin). Beyond 78 arcmin, the surface brightness drops very abruptly, such that we no longer detect any significant cluster emission in the last annulus (78--88 arcmin). However, we can place an interesting upper limit on the emission measure in this region, more than a factor of 10 below that in the preceding annulus. This is consistent with the presence of a shock. The 2$\sigma$ upper limit plotted in Fig. \ref{projsw} was obtained assuming the temperature in that annulus to be 4~keV, which is approximately the value in the outermost regions where this quantity is constrained. If this is indeed a shock, one would expect the temperature to be lower than 4~keV, which would only tighten the reported upper limit (for a gas with a temperature of 0.7 keV, the upper limit decreases by about 30\%).

\subsubsection{Deprojection Analysis}

\begin{figure}
\begin{center}
\includegraphics[width=\columnwidth]{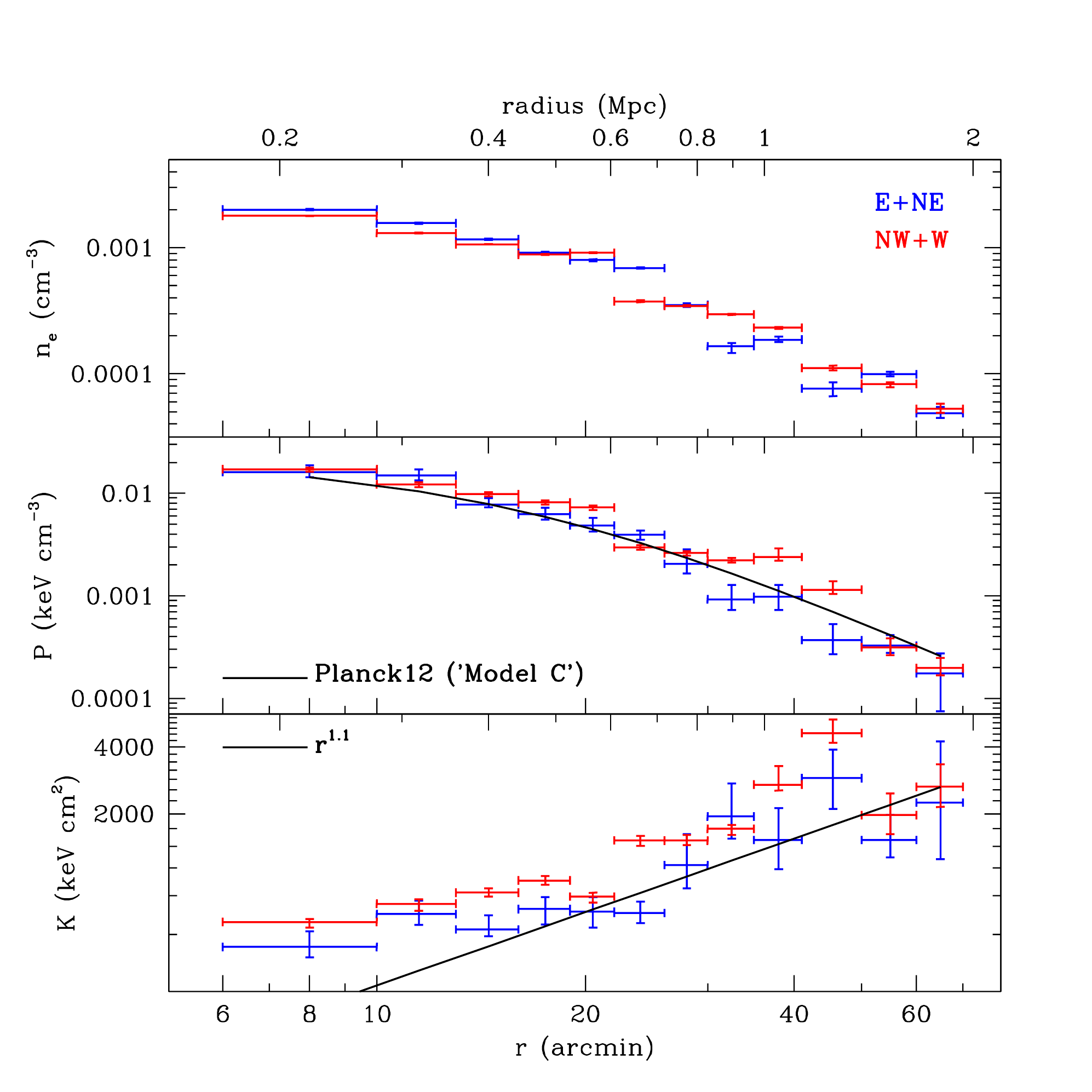}
\end{center}
\caption{Deprojected radial profiles of the electron density, entropy, and pressure obtained from the Suzaku data along the more relaxed NW+W and E+NE directions. \label{deproj}}
\end{figure}

We have produced deprojected profiles of the electron density, entropy, and pressure, along two different directions. Along one direction, we combined the E and NE spectra, while along the second direction we grouped the W and NW arms. By construction of the model, we necessarily rely on the assumption of spherical symmetry, which is likely to be only a very rough approximation for an unrelaxed system like Coma - however, the similarity of the projected profiles in Figure \ref{projrel} argues that this is still a sensible assumption. The results are shown in Figure \ref{deproj}. The errors in this case have been derived using a Markov-Chain Monte Carlo (MCMC) analysis.
 
\cite{PlanckComa} estimate the scale radius of the Coma Cluster to be $r_{500} \approx (47 \pm 1)$ arcmin by using the existing XMM-Newton mosaic to iteratively calculate the $Y_X = M_{gas}\times kT$ parameter of the cluster and using the $Y_X - M_{500}$ scaling relation calibrated from hydrostatic mass estimates in a nearby cluster sample \citep{arnaud2010}. We have thus chosen to scale our radii in Figure \ref{deproj} by this value of $r_{500}$, which would imply $r_{200}\sim70\:{\rm arcmin}\sim2$ Mpc. 

The electron density profile towards the E shows two marked discontinuities at 30 and 40 arcmin, which reflect the surface brightness profile jumps discussed in the previous section (see Fig.~\ref{sbprof}). There is a clear excess pressure along the NW/W compared to the E/NE, centered around a radius of 40 arcmin.  


In Figure \ref{deproj}, we compare our measurements of the ICM pressure with the SZ results presented by \cite{PlanckComa}. On average, there is a good agreement between the Suzaku measurements and the best fit model to the azimuthally averaged Planck data (Model C in \citealt{PlanckComa}), although around 40 arcmin, the E/NE pressure is systematically lower and the NW/W systematically higher than this model. 
Given that the data used to determine the Planck model represents an average over all azimuths, the agreement at large radii, where Suzaku only covers a fraction of the available volume, is remarkable.

Standard large scale structure formation models predict that, in the absence of other sources of heating or cooling except shock heating as gas accretes onto clusters, the gas entropy $K$ should behave as a power-law with radius, which can be written as
\begin{equation}\label{eqnK}
\frac{K}{K_{500}}=1.47\left( \frac{r}{r_{500}} \right)^{1.1}
\end{equation}
where $K_{500}=106 {\rm\:keV\:cm^2}(M_{500}/10^{14}M_\odot)^{2/3}f_b^{-2/3}E(z)^{-2/3}$ (\citealt{Pratt10}, adapted from \citealt{voit2005MNRAS}), and we have assumed the baryon fraction at $r_{500}$ to be $f_b=0.15$. This expected power-law behavior is overplotted together with the measured entropy profiles in the bottom panel of Figure \ref{deproj}. We see a clear entropy excess out to 1.06$r_{500}$ (except potentially for the dip in entropy along the E between 10--30 arcmin, which is however likely associated with an infalling subgroup, as will be discussed in Section \ref{disc}). This excess likely reflects the violent merging events experienced by the Coma Cluster. Beyond 1.06$r_{500}$, however, the entropy is consistent with the expected trend, although the statistical uncertainties are large.

\section{Systematic Uncertainties}

Here, we describe in detail the systematic uncertainties related to background subtraction, solar wind charge exchange, and stray light.

\subsection{Background Fluctuations}\label{sectbkgsys}

\begin{figure*}
\begin{center}
\includegraphics[width=0.49\textwidth]{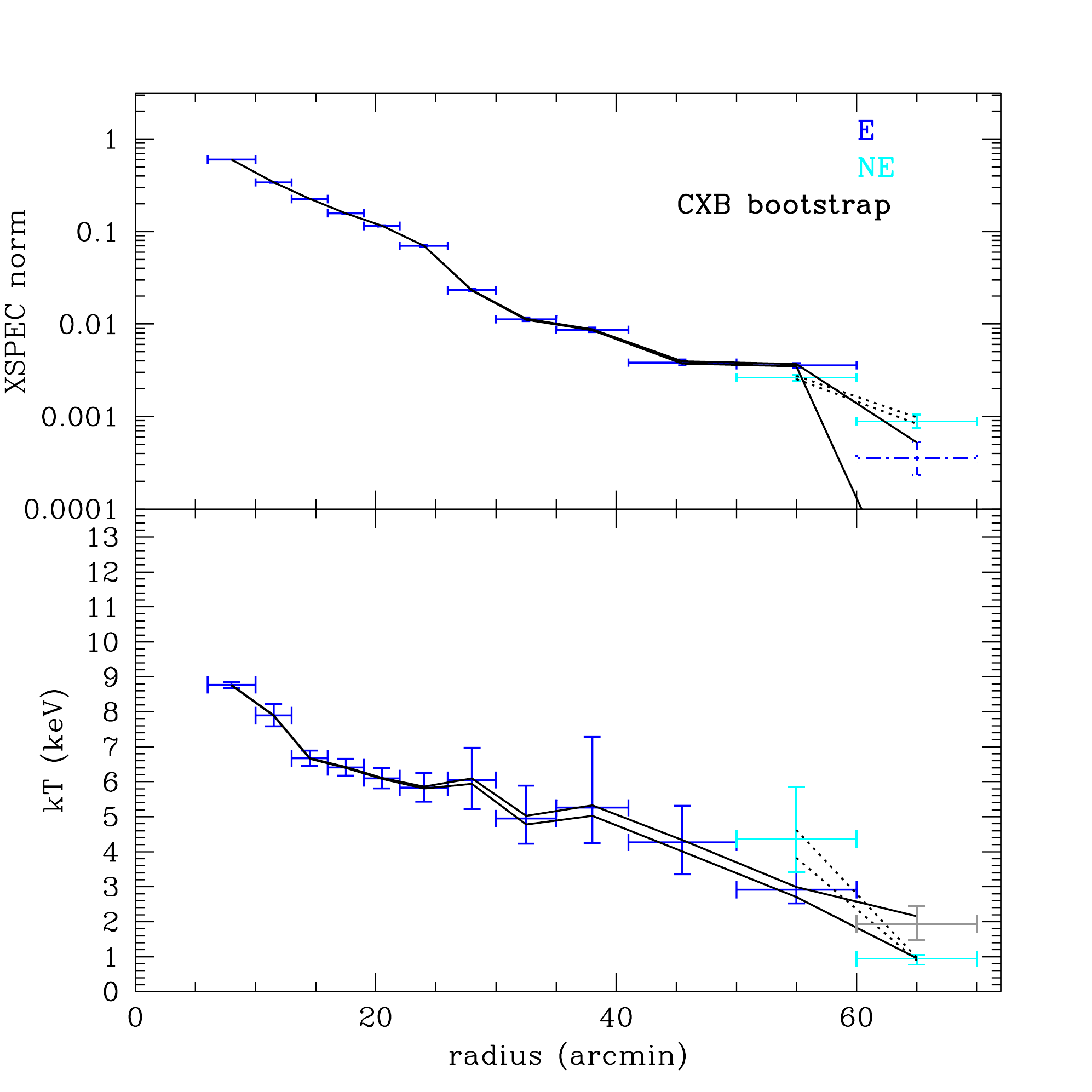}
\includegraphics[width=0.49\textwidth]{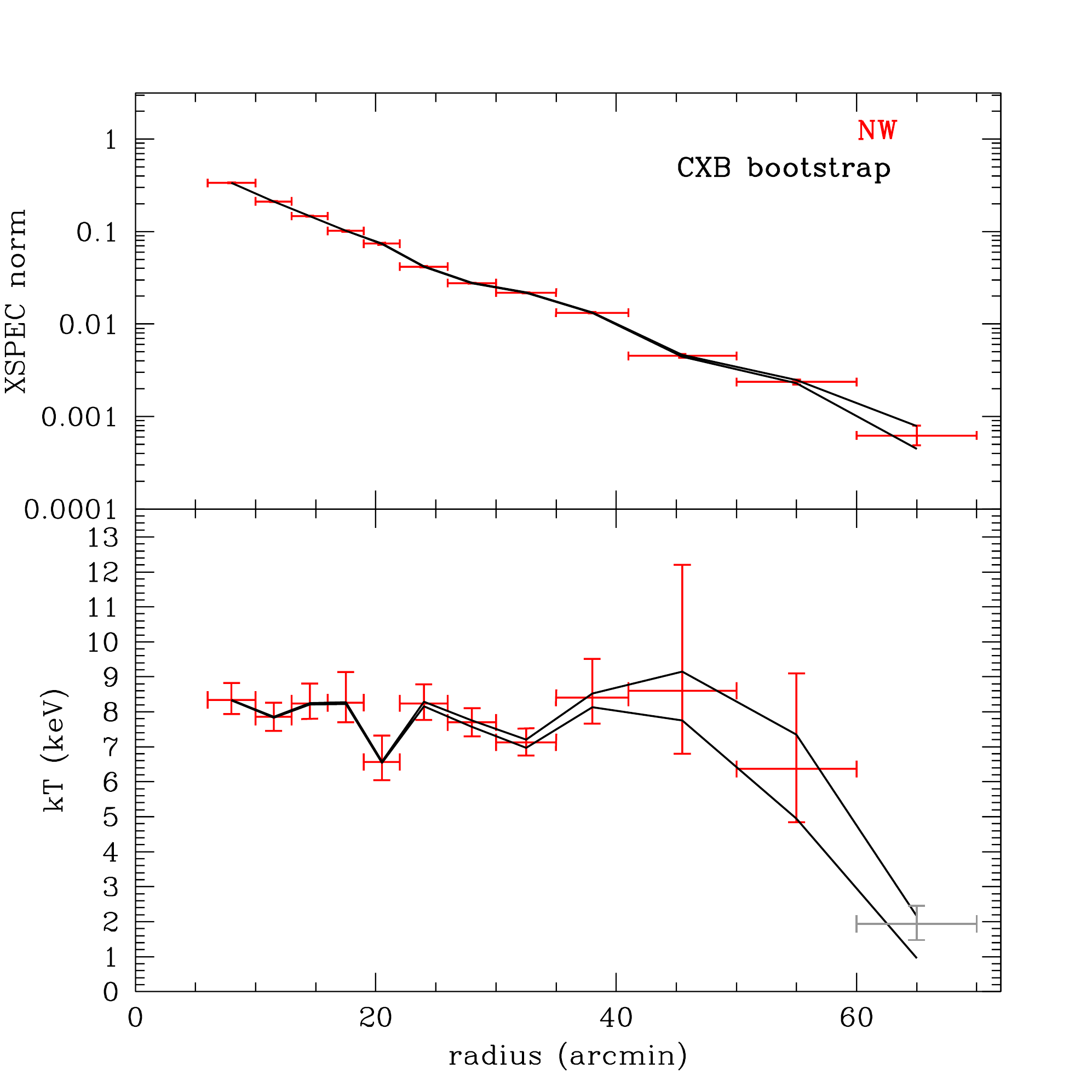}
\includegraphics[width=0.49\textwidth]{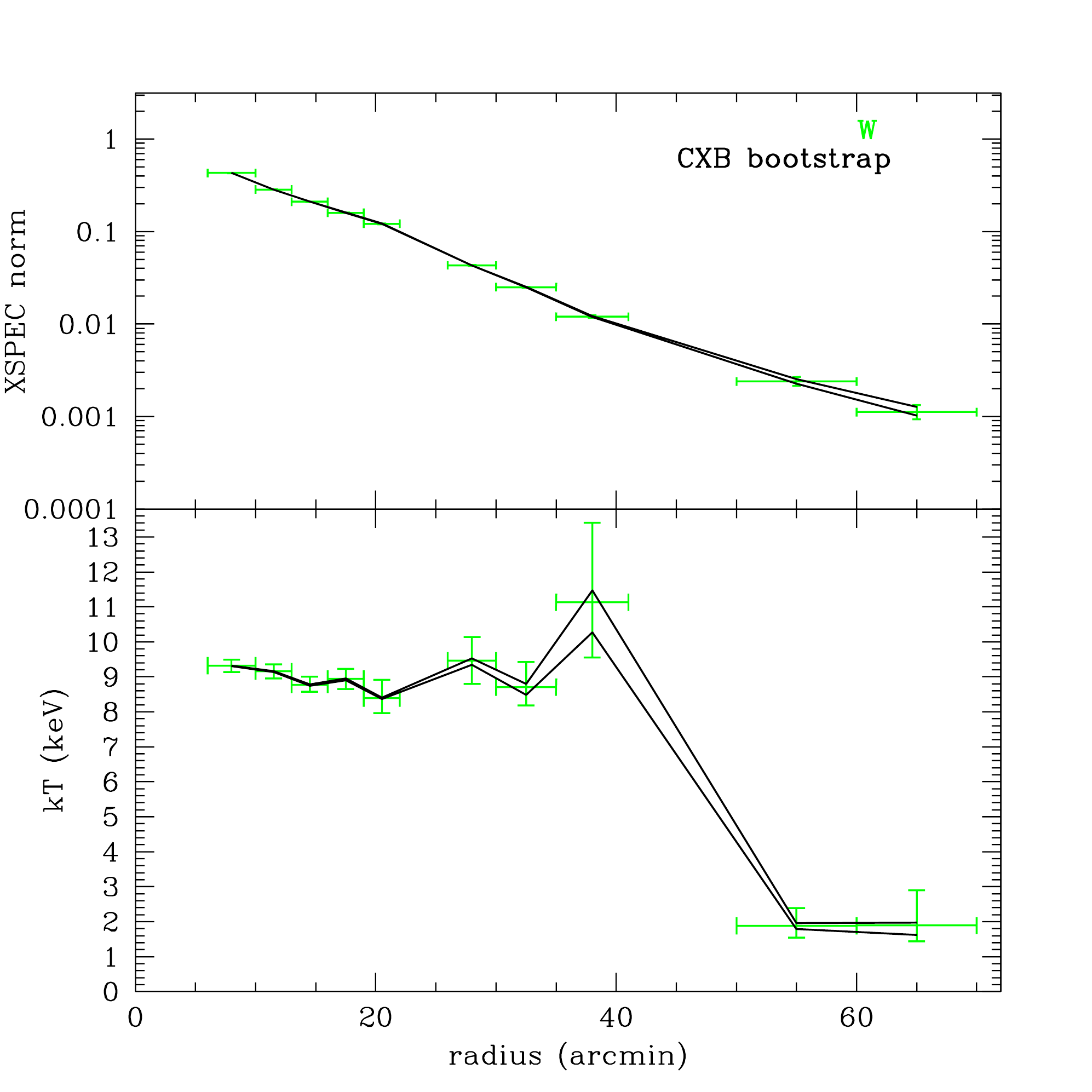}
\includegraphics[width=0.49\textwidth]{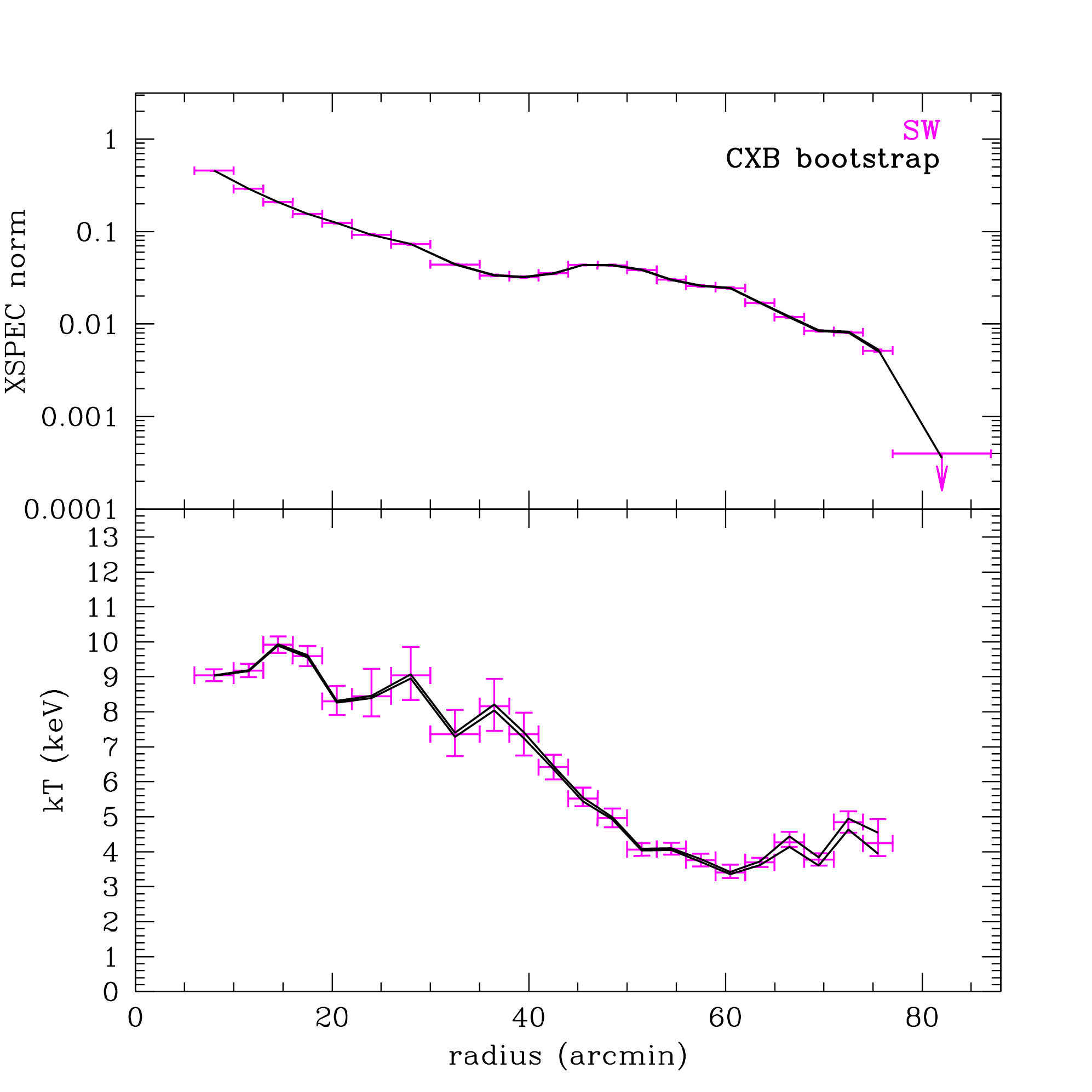}
\end{center}
\caption{Systematic uncertainties due to the CXB variance estimated using bootstrap analysis.}\label{sys}
\end{figure*}

In order to estimate the uncertainty on our measurements due to CXB fluctuations, we divided the Suzaku data used to obtain the background model described in Section \ref{cxb_default} into spectra 
with approximately equal exposure times and sizes of the extraction region. Each pointing was divided into two spatial regions, and the pointings with longer exposure time such as ne6 were split into several event files, resulting in a set of 26 independent spectra for each detector. Following a bootstrap approach, we randomly picked 26 out of these 26 background realizations, allowing repetitions, and re-determined the best-fit background model in the same way as the default background model (fitting first the CXB power-law index in the hard band, and then all the other parameters using the full band). We then re-fit each spectral region associated with cluster signal with this new background model. We repeated this procedure 1000 times, each time randomly picking a different set of the 26 background realizations, and determined the distribution of best-fit cluster temperatures and spectral normalizations obtained with these different CXB models, and the corresponding 68\% confidence intervals. The results are presented in Figure \ref{sys}. Since the temperature was not constrained in the last annulus towards the NW and E, in the corresponding panels of Figure \ref{sys} we show in gray the common average temperature among all 4 relatively relaxed azimuths for this annulus.

The uncertainties due to CXB fluctuations are always smaller than the statistical errorbars, with the exception of the last data point along the E arm, where the spectrum normalization is unconstrained in the bootstrap analysis. The upper limit on the normalization in the last annulus towards the SW is also robust.

We note that the results reported here differ somewhat from those obtained by \cite{Akamatsu13}, who analyzed the SW arm of the mosaic and report a detection of X-ray emission out to larger radii than in this work (approximately 96 arcmin). This difference is due to the choice of observations used to estimate the CXB. \cite{Akamatsu13} determined their CXB model parameters using a single offset Suzaku observation, COMA BGK (Observation ID = 802083010), which was centered at a much larger distance from the observed regions (6 degrees from NGC 4839, which is far outside the image shown in Fig. \ref{coma_image}). Using the CXB parameters reported in \cite{Akamatsu13}, we find significant detections above this model in all our background spectra. However, the emissivity does not show a decreasing trend with radius, suggesting that this emission is not cluster-related, but most likely due to an elevated Galactic foreground which is not present in the COMA BGK region located further away from the cluster center.

\subsection{Solar Wind Charge Exchange}

Charge exchange between heavy ions in the solar wind and neutral atoms in the Earth's geocorona and in the solar magnetosphere produces emission lines of highly-ionized C, N, O, Ne, and Mg which can act as an additional foreground to the cluster emission. The geocoronal solar wind charge exchange (SWCX) flux can vary on timescales of seconds \citep[e.g.][]{Fujimoto07}. Empirically, when the solar wind proton flux is below $4\times10^8$ cm$^{-2}$ s$^{-1}$, Suzaku spectra do not show strong SWCX signatures \citep{Fujimoto07,Yoshino09}.

We calculated the solar-wind proton flux at the Earth using WIND SWE (Solar Wind Experiment\footnote{http://web.mit.edu/afs/athena/org/s/space/www/wind.html}) data and find that the proton flux reaches a level above $4\times10^8$ cm$^{-2}$ s$^{-1}$ during the observation dates of fields e1, e2, e3, nw5, nw65, w1, w4, w5, w6, sw4. The maximum proton flux during these observations is $8\times10^8$ cm$^{-2}$ s$^{-1}$, which means we are dealing, if at all, with mild flares (during strong flares, the proton flux can reach levels as much as an order of magnitude higher). The spectra of nw65 and w5 do not show anomalous features, or an anomalous soft flux, compared to the seven Suzaku background pointings during which the proton flux was low; moreover, the best-fit spectral parameters for the cluster emission agree well between the e3 and e35 pointings (the latter also shows a very low and stable proton flux). Therefore we do not expect our results to be significantly influenced by SWCX flares.

\subsection{Stray Light}

The effects of stray light contamination (light scattered from very bright sources outside the field of view, in particular the cluster center) are mitigated by the choice of target -- Coma is a non-cool core cluster, whose central surface brightness is an order of magnitude lower than in the Perseus Cluster, which has also been observed extensively with Suzaku. We have found that stray light does not hinder robust measurements of the temperature and density in the outskirts of the Perseus Cluster (Urban et al., in prep.), and thus do not expect significant effects in the case of the Coma Cluster where this effect should be lower. 

This is supported by the fact that the best-fit spectral parameters for the cluster emission agree well between the e3 and e35 pointings -- because of the mirror geometry, we expect that the two different roll angles of the 
e3 and e35 pointings would produce very different levels of stray light contamination, thus if stray light had an important effect, we should measure different parameters from the two observations.

Off-axis observations of the Crab nebula have shown that, for pointings centered more than 80 arcmin away, the interior of a circle centered on the bright source passing through the boresight is free from any stray light. In other words, for a pointing centered beyond 80 arcmin from the cluster core, the half of the field of view which is closer to the cluster center is stray-light free, while the half located further away may be influenced by stray light. We have fitted each of these halves of all the pointings at $r>80$ arcmin towards the E, NE, W, and NW with the default background model described in Sect. \ref{cxb_default}, allowing for an excess thermal component representing possible stray light. We find a significant excess in the halves of the fields of view of nw55, nw6, and nw7 located further from the Coma Cluster core. None of these 3 pointings show evidence of SWCX contamination (see previous subsection). We also find a significant excess in the near-half of w6, which is likely associated with the background group located here, but no excess in the far halves of w5 or w6. In conclusion, the E, NE, and W directions do not show any evidence of stray light influence; however, the stray light pattern is not symmetric to a 90 degree rotation of the field of view, and it is possible that the NW arm observations were performed in a configuration with higher stray light. Fortunately, however, unlike most NW arm pointings, the observations nw2 and nw3.5 have roll angles known to minimize stray light (with the diagonal of the field of view pointing towards the bright cluster center). These observations cover annuli located between  22--41 arcmin and 41--60 arcmin, respectively. We have verified that, in regions where nw3 and nw3.5 overlap, including the nw3 spectra (which would be more affected by stray light due to the mirror geometry) does not significantly change the best-fit values of the temperature and normalization. Thus, the only region left which may be influenced by stray light is the last annulus towards the NW, covering radii from 60--70 arcmin. Since the emission measure here does not appear particularly high compared to other azimuths, the effect of stray light should be small.

We have verified that the best-fit cluster parameters presented here are not affected either by excluding the halves of nw55, nw6, and nw7 affected by stray light from the background modeling, or by including the halves of w5 and w6 which show no signal above the nominal background.

\subsection{Comparison with ROSAT}


Due to the low and stable background and larger field of view of ROSAT, it is sensible to compare the surface brightness profiles obtained with the PSPC to those obtained from Suzaku. This can help, on one hand, to rule out systematic issues related to stray light and background subtraction uncertainties and, on the other hand, to test how representative the surface brightness probed by the Suzaku mosaic is compared to a more complete coverage of the outskirts available with ROSAT.

\begin{figure}
\begin{center}
\includegraphics[width=1.1\columnwidth,bb=18 0 592 374]{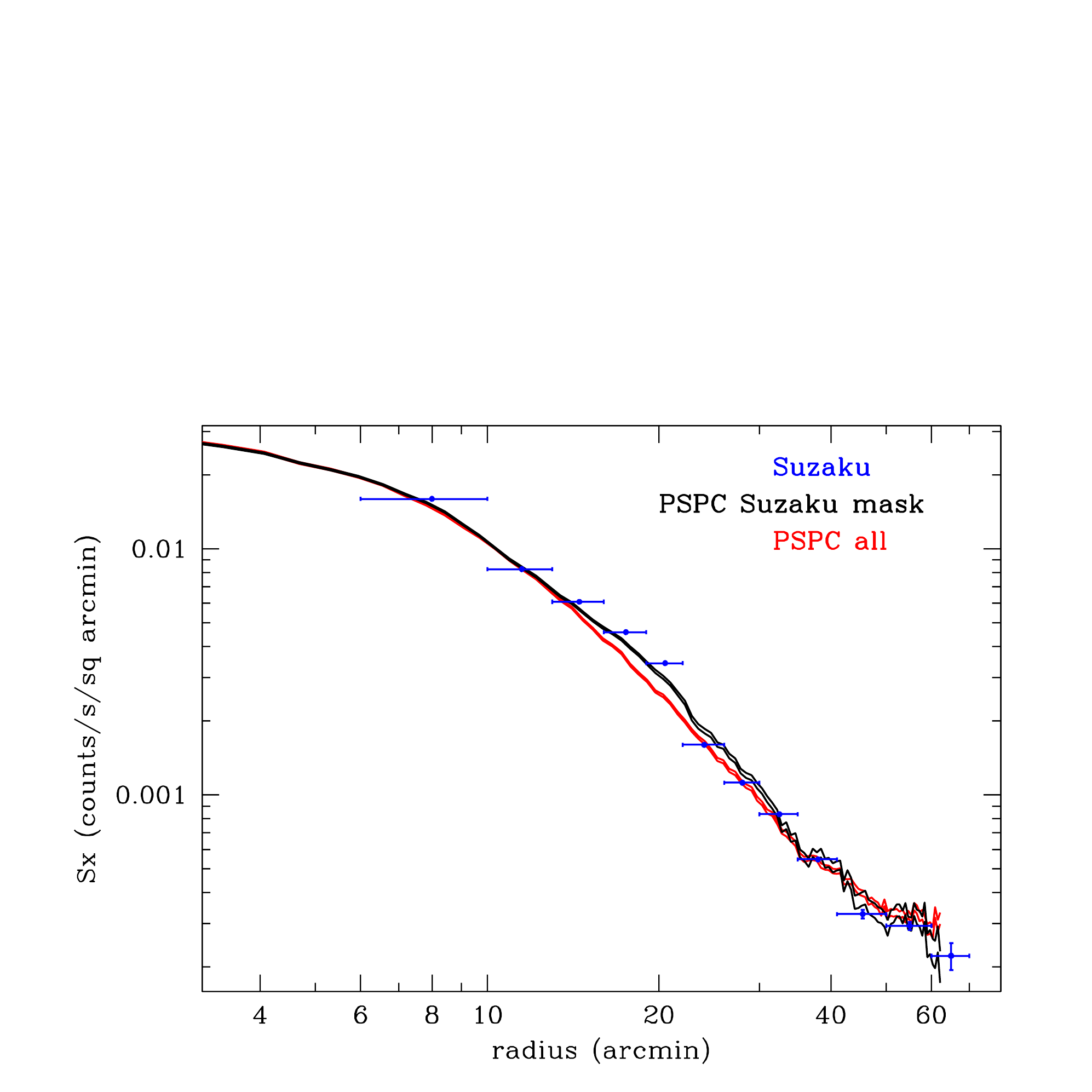}
\end{center}
\caption{Surface brightness profiles in the 0.7--2 keV band obtained with ROSAT. The two black lines show the 1$\sigma$ confidence interval for the surface brightness profile obtained using only the regions in the PSPC mosaic which also had Suzaku coverage; in red, we show the same using all the relatively undisturbed azimuths in the ROSAT image. For comparison, in blue we show the predicted ROSAT counts based on the best-fit Suzaku temperature and normalization profiles. The direction of the SW subcluster was excluded in all cases. 
\label{pspcfak}}
\end{figure}

We used the ROSAT-PSPC response to predict the count rate corresponding to the best-fit model obtained with Suzaku in each annulus, and compared this to the observed 0.7--2 keV ROSAT surface brightness profile obtained in two different ways. In one case, we used only the regions in the PSPC mosaic which also had Suzaku coverage, excluding the SW arm; in the other case we obtained the surface brightness profile for all azimuths in the ROSAT image, except for a 60-degree wedge around the SW subcluster. The profiles are shown in Figure \ref{pspcfak}. In general, we notice a good agreement between all three profiles. Between about 15--30 arcmin, the Suzaku mosaic preferentially probes regions of higher surface brightness seen towards both the E and W. However beyond this radius, there is no indication that the surface brightness probed with Suzaku is atypical of the average measured from the PSPC mosaic. 

\section{Discussion}\label{disc}

\subsection{Complex morphology of a merging cluster}
\begin{figure}
\begin{center}
\includegraphics[width=\columnwidth]{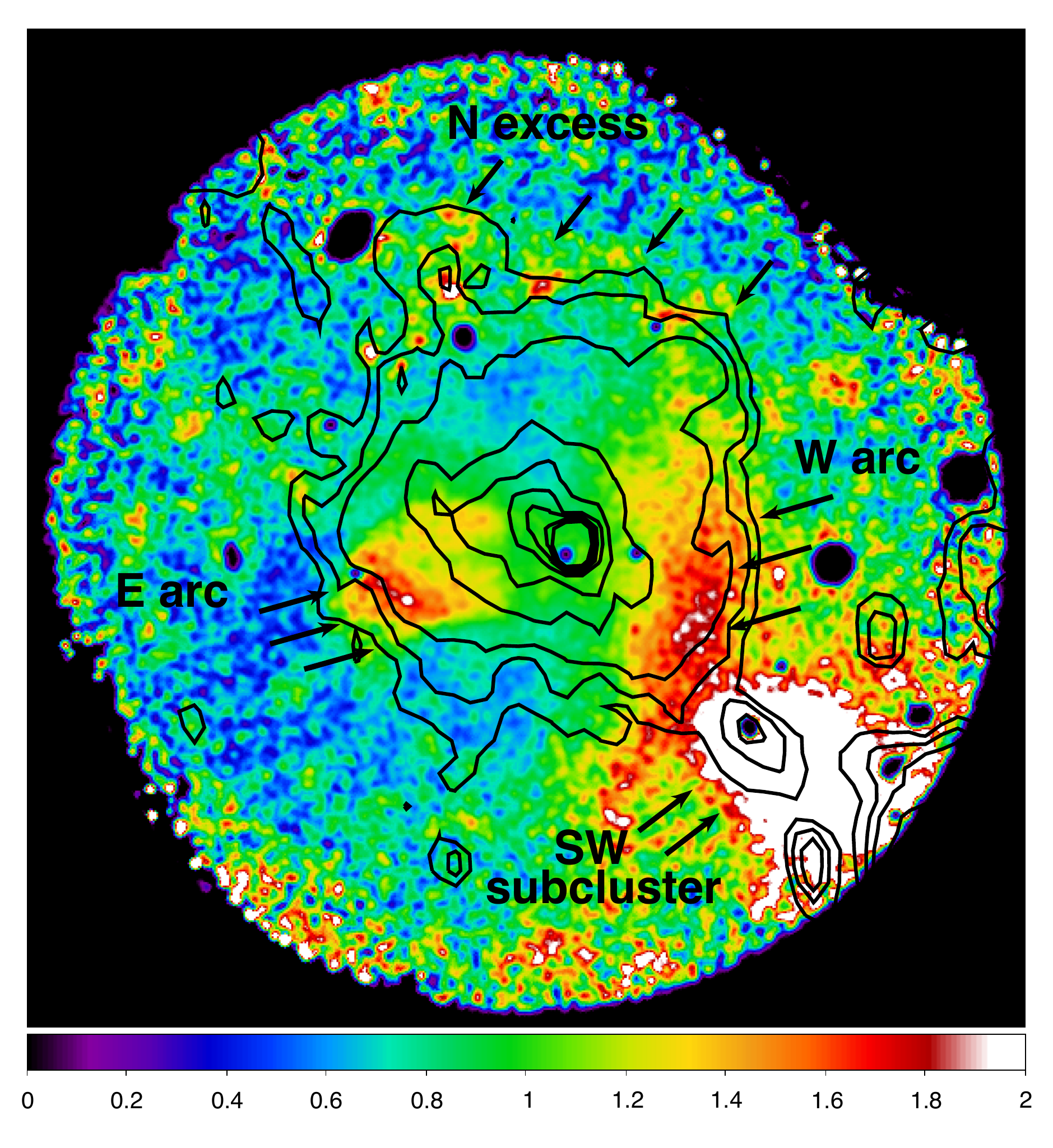}
\end{center}
\caption{ROSAT image divided by the azimuthally averaged profile. 352 MHz WSRT radio contours from \cite{Brown11} are shown in black. \label{coma_rat}}
\end{figure}

The Coma Cluster is currently undergoing a merger with a substructure located towards the southwest from the cluster center. Based on XMM-Newton data, \cite{neumann2001} argue that the southwestern subcluster is experiencing its first infall. The main cluster itself is a non-cool core system featuring two central galaxies with similar brightness, rather than a single dominant brightest cluster galaxy (BCG). Therefore, it is assumed that its formation history has included at least one other relatively recent major merger. This is confirmed by observations in the radio band, which show that the Coma Cluster hosts both a giant radio halo \citep[Coma C,][]{Willson70} associated with merging activity, and a peripheral radio relic \citep[1253+275,][]{Jaffe79,Giovannini85} in the direction of the infalling south-western subcluster. 

In Figure \ref{coma_rat}, we show the result of dividing the ROSAT-PSPC image by the azimuthally averaged ROSAT surface brightness profile. Apart from the southwestern infalling subcluster, we note the presence of two arc-shaped regions with excess surface brightness towards the east and west of the main cluster core. These features were previously discussed in detail by \cite{neumann2003} based on XMM-Newton data.
The presence of these structures is also reflected by the Suzaku surface brightness profiles, which show strong gradients between 25--35 arcmin along the E and W arms. The outer edge of both of these X-ray bright regions coincides well with the outer edge of the giant radio halo observed at 352 MHz with the Westerbork Synthesis Radio Telescope (WSRT) \citep{Brown11}. In addition, we note here an arc of enhanced surface brightness which seems to track the entire northern edge of the giant radio halo, although at a much smaller contrast than the E and W features.

The radial entropy distribution toward the E shows a dip around r$\sim25$ arcmin. Within the statistical uncertainties of our measurement, the temperature profile appears constant across this feature. Therefore, the dip in the entropy profile appears to be mainly due to the enhanced density, and the feature has an excess pressure with respect to the gas at larger radii. \citet{PlanckComa} arrive at a similar conclusion, finding a steepening of the SZ pressure profile towards the SE from the cluster core which is associated with sharp features in both the X-ray and radio brightness, but does not show an evident jump in the temperature profile. With respect to our choice of cluster center coordinates, this feature in the Planck data is seen at a radius of 35 arcmin, at azimuths of 195--240$^\circ$, which are just south of the coverage of our Suzaku mosaic. It is therefore possible that our X-ray observations do not probe the maximum pressure jump.

The excess X-ray emission and overall lower temperature towards the E are most likely due to the infall of a small group centered around the two galaxies NGC 4921 and NGC 4911, as argued by \cite{vikhlinin1997} and \cite{neumann2003}. These galaxies are located at the same position as the E arc in Fig. \ref{coma_rat}. Since the eastern arc is over-pressured, the infalling group is probably still moving supersonically and should drive shocks into the surrounding ICM, which likely explains the second surface brightness edge seen at a larger radius of 40 arcmin towards the E. 

The NW profile shows a surface brightness jump at 40 arcmin, corresponding to the edge of the radio halo, which extends furthest in radius along this direction. At the same radius, the deprojected thermodynamic profiles  show a sharp drop in density and pressure. In particular, the NW/W pressure profile deviates from the average pressure profile measured from Planck, showing a clear excess.

Along the western direction, \citet{PlanckComa} found a marked local steepening of the radial gradient of the SZ signal measured with Planck at a radius of about 30 arcmin, and interpret this feature as a shock with a Mach number of about two. They also report jumps in the X-ray brightness, temperature, and radio brightness associated with this feature. While, at the same radius, we observe an X-ray surface brightness excess in the form of the western arc in Fig. \ref{coma_rat}, whose edge coincides with that of the 352 MHz radio halo, the X-ray temperature profile along this direction does not drop at 30 arcmin, but is consistent with the presence of a discontinuity further out, beyond a radius of 40 arcmin. The most probable reason why \citet{PlanckComa} report a temperature jump at 30 arcmin that is not seen in our data is the difference between azimuthal coverage of the extraction regions being used. \citet{PlanckComa} use azimuths that run W-SW, as opposed to W-NW in our case. The presence of this azimuthal asymmetry was noted already by \citet{Brown11}, who found a temperature decrease across the southern half of the western radio halo edge, but an apparent increase along the northern half. This difference in itself is noteworthy, and may be related to the hint of shock heating seen at a larger radius of 40 arcmin towards the NW. 

While we do not detect significant temperature jumps in the shapes of the temperature profiles along individual arms, the azimuthally averaged temperature profile does seem to drop sharply beyond a radius of about 40 arcmin. 

We are therefore presented with a complex picture of the Coma Cluster ICM, with multiple edges which all appear to be over-pressured. Towards the E and W, the inner features, or ``arcs'', at radii of about 30 arcmin, which are over-pressured due to a higher gas density but not due to a temperature jump (at least, for our azimuthal coverage), trace well the edge of the 352 MHz radio emission along these directions. Beyond these features, at a larger radius of 40 arcmin, we find a steep temperature gradient which may represent an outward propagating shock front, possibly surrounding the cluster center at all azimuths. This feature is associated both with a temperature and a density jump, and its radius corresponds to the largest extent of the spherically asymmetric giant radio halo (which stretches out furthest towards the N). 

Merger-driven shocks and turbulence may lead to the amplification of the magnetic fields \citep[e.g.][]{dolag2002} and to (re)-acceleration of electrons \citep[e.g.][]{ensslin1998}, giving rise to radio emission. The observed spatial coincidence between the observed jumps in the pressure profiles and the outer edge of the giant radio halo lend further support to the picture that supersonic large-scale motions produced by major mergers lead to the origin of giant radio halos. 
 
Along the direction of the ongoing SW merger, the ICM temperature shows a marked decline starting at 40 arcmin, the same radius where we found a drop in the azimuthally averaged temperature profile along the relaxed directions. This steep gradient is found in the radial range of 40--55 arcmin, beyond which the temperature remains practically constant over one Mpc in radius (from about 55--78 arcmin). This could argue that the infalling southwestern subcluster generates mixing in the ICM at large radii which causes the temperature to homogenize over this wide radial range. 

In the last annulus (78--88 arcmin), the spectrum normalization drops dramatically. We obtain a robust lower limit on the surface brightness jump of a factor of 13, which is indicative of the presence of a shock, or else a steep gradient unresolved by Suzaku. The presence of a shock at this location could explain the powering of the SW radio relic through particle re-acceleration. Hints for the presence of this shock have also been reported from XMM-Newton data \citep{Ogrean13} and from Suzaku data \citep{Akamatsu13} (but see Section \ref{sectbkgsys}). 

\subsection{Entropy at large radii}

Standard large scale structure formation models show that matter is shock heated as it falls into clusters under the pull of gravity. Simple theoretical models of this process predict that the entropy $K$ should behave as a power-law with radius, according to Equation \ref{eqnK}. However, as more and more measurements of the thermodynamic profiles of the ICM at large radii are being performed, primarily with the Suzaku satellite, a flattening of the entropy profile away from this expected behavior near the virial radius has been observed in many systems.

\citet{Eckert13} have recently argued that the entropy excess in cluster cores may have caused the normalization of the $K\propto r^{1.1}$ model to be overestimated in previous publications, where this normalization was allowed as a free parameter in the fit, rather than being fixed based on Eqn. \ref{eqnK}. However, \citet{Walker13} show that, even when fixing both the normalization and index of the power-law model for the expected entropy behavior, there is still an entropy deficit at large radii in most systems for which thermodynamic measurements with Suzaku have been published (which were so far predominantly relaxed clusters, as mentioned in Section \ref{intro}).

In the case of the Coma Cluster, there is a very clear entropy excess above the expected profile from Eqn. \ref{eqnK} inside of $r_{500}$. This reflects the violent merging events whose traces are seen in the morphological features described in the previous subsection. Due to this large excess, the shape of the entropy profiles appears to flatten at larger radii. However, we find that, beyond $r_{500}$, the entropy profile is consistent with Eqn. \ref{eqnK}, within error bars. Therefore, even though the shape of the Coma entropy profiles appears similar to that seen in more relaxed systems, the overall normalization is quite different. The scaled entropy values at corresponding scaled radii in the outskirts are thus larger than in most cool core clusters observed with Suzaku albeit, unfortunately, due to the large error bars on the Coma measurements, this difference is only marginally significant.

If there is indeed a systematic difference between the entropy levels in the outskirts of cool core and non-cool core clusters, this could be explained in several ways. For example, \citet{lapi2010} and \citet{cavaliere2011} propose a weakening of the strength of the accretion shock as clusters get older. This would predominantly affect the outskirts of relaxed, cool-core clusters, and is consistent with the measured scaled entropy being larger in the outskirts of non-cool core clusters like Coma. 
Alternatively, the entropy deficit in cluster outskirts may be due to gas clumping at large radii \citep[e.g.][]{SimionescuSci}, causing the measurements of the gas density obtained from the bremsstrahlung emission to be biased high. In this case, the fact that there is no evidence for such an entropy deficit in Coma -- and no evidence for a pressure excess above the SZ measurements made by Planck, which would also result from a bias in the gas density -- suggests that gas clumps may be more easily destroyed in a dynamically active cluster.
\cite{Hoshino10} and \cite{Akamatsu11} on the other hand suggest that the entropy deficit in the outskirts may be due to the electron temperature being biased low because classical shock fronts heat the ions practically instantaneously, whereas electrons equilibrate on a collisional timescale. We would, however, expect this effect to be much stronger in a dynamically active system \citep{Rudd09}, which is contrary to the scaled entropy being higher in Coma than in many cool-core clusters. The electron-ion nonequilibrium therefore is unlikely to play a large role in influencing the entropy of the ICM.

Azimuthal variations are also very important for understanding the dynamics of cluster outskirts. The boosted surface brightness (and thus density) associated with the infalling subcluster towards the SW would significantly lower the azimuthally averaged entropy profile of the Coma Cluster, causing a deficit at large radii compared to the expectations from Eqn. \ref{eqnK}. In the only other clear mergers studied to date out to $r_{200}$, A2142 and A3667, a flattening or even inversion of the entropy profile at large radii has been reported \citep{Akamatsu11,Akamatsu12}. However, for both of these systems, only the direction along the most recent major merger axis has been studied, and it is currently unclear whether the entropy along relatively less disturbed directions would be in better agreement with the predictions, as in the case of the Coma Cluster. Moreover, while the best-fit scaled entropy in the outermost annulus in Coma, $K/K_{500}$, is about 1.6 times higher than the average immediately inside $r_{200}$ of the Perseus Cluster (Urban et al., in prep), the latter system exhibits significant azimuthal scatter as well, with the entropy along the major axes following Eqn. \ref{eqnK} more closely than along the minor axes. It appears thus that the entropy profile in cluster outskirts may be equally sensitive (or even more sensitive) to the large-scale environment, which dictates azimuthal asymmetries at large radii, than to the average dynamical state of the cluster.

\section{Conclusions}

We have presented results from a Suzaku Large Project targeted on the Coma Cluster, which is the nearest and X-ray brightest hot ($\sim$8~keV), dynamically active, non-cool core system, focusing on the thermodynamic properties of the ICM on large scales.

Along the relatively undisturbed directions covered by the Suzaku mosaic, namely the E, NE, NW, and W, we detect emission from the ICM of the Coma Cluster out to a radius of up to 70 arcmin (2~Mpc). Both the temperature and spectral normalization are in broad agreement and show consistent radial trends along those directions, with the temperature decreasing from about 8.5~keV at the cluster center to about 2~keV in the outskirts. The temperature along the eastern arm is systematically lower than towards the NW and W from the center of the cluster out to 50 arcmin (1.4 Mpc). 
Towards the SW, an ongoing merger has a major impact on the surface brightness distribution, which is significantly boosted compared to the relatively undisturbed azimuths over a wide range of radii, from about 25 arcmin (700 kpc) to the edge of the detection limit. As a result of this boost, we detect X-ray emission out to larger radii than towards the other directions (up to $\sim$2.2 Mpc). 

Apart from the southwestern infalling subcluster, the surface brightness profiles show multiple edges around radii of 30--40 arcmin (840--1120 kpc). These regions appear over-pressured and they most likely originate from merger induced large scale supersonic gas motions, which are driving shocks into the ICM. While we do not detect significant temperature jumps in the shapes of the eastern or western temperature profiles alone, the azimuthally averaged temperature profile, as well as the deprojected density and pressure profiles, all show a sharp drop consistent with an outward propagating shock front located at 40 arcmin, corresponding to the outermost edge of the giant radio halo observed at 352 MHz with the WSRT \citep{Brown11}. The shock front may be powering this radio emission. The surface brightness drops dramatically, by a factor of at least 13, between 75--82 arcmin (2.1--2.3 Mpc) along the direction of the infalling southwestern subcluster, indicating the presence of an additional shock likely related to the SW radio relic seen at this location.

Beyond $r_{500}$, the entropy profiles of the Coma Cluster along the relatively relaxed directions are consistent with the power-law behavior expected from simple models of gravitational large-scale structure formation. The pressure is also in agreement at these radii with the expected values measured from SZ data by \citet{PlanckComa}. The scaled entropy values at corresponding scaled radii in the outskirts are smaller in most cool core clusters observed with Suzaku. 
This would suggest either that the gas clumping factor is smaller in the outskirts of the Coma Cluster than in cool core clusters, or that the accretion shocks weaken as cool core clusters become more relaxed, but do not weaken for more active systems like Coma. 
However, due to the large statistical uncertainties associated with the Coma Cluster measurements, we cannot exclude an entropy flattening in this system consistent with that seen in cool core clusters. Moreover, the entropy profile in cluster outskirts does not depend only on the dynamical state of the cluster, but also on the large-scale environment, which dictates azimuthal asymmetries at large radii.


\acknowledgments

We thank our referee, P. Mazzotta, for his constructive comments and L. Rudnick for providing the 352 MHz WSRT map.
Support for this work was provided by NASA through Einstein Postdoctoral Fellowship grant number PF9-00070 awarded by the Chandra X-ray Center, which is operated by the Smithsonian Astrophysical Observatory for NASA under contract NAS8-03060. 
We further acknowledge support from awards GO2-13145X and NNX12AE05G.
The work was supported in part by the U.S. Department of Energy under contract number DE-AC02-76SF00515.
The authors thank the Suzaku operation team and Guest Observer Facility, supported by JAXA and NASA. 

{\it Facilities:} \facility{Suzaku}.

\bibliography{bibliography,clustersnewest,clustersvirial}

\bibliographystyle{apj}

\appendix

\section{The list of Suzaku observations}

\begin{table*}[!h]
  \caption{Suzaku observations used in this work.}
  \label{tab:observations}
  \scriptsize
  \centering
  \begin{tabular}{lccccccc}
  \hline
    Target & ObsID & Obs. date & RA & Dec & Exposure  & $n_H$ \\
     & &  &  &  &(ks) &(10$^{20}$ cm$^{-2}$) \\  
    \hline
E1     & ae806030010  & 2011-06-16  &   13 00 54.91 & +27 54 06.8 & 9  & 0.78  \\             
E2       &  ae806031010   & 2011-06-17 & 13 02 07.06 & +27 53 47.8 & 8 & 0.80 \\            
E3        & ae806032010    & 2011-06-17 & 13 03 18.67 & +27 53 45.6 & 5 & 0.80  \\     
E3.5    & ae806032020  & 2011-12-22  &  13 03 43.20 &	+27 54 27.4  & 9 & 0.79 \\            
E4          & ae806033010   &  2011-06-17 & 13 04 31.01 & +27 53 40.9 & 19 & 0.78  \\               
E5         & ae806034010   &  2011-06-18 & 13 05 42.86 & +27 53 25.4 & 13 & 0.99  \\               
E5.5      &  ae806035010  &  2011-06-18 & 13 06 26.76 & +27 53 39.8 & 11 & 0.99  \\               
E6       & ae806036010 &   2011-06-19   & 13 07 05.30 & +27 53 12.1 & 10  & 0.99 \\               
E6.5    & ae806037010  &   2011-06-19   & 13 07 48.62 & +27 53 08.5 & 12 & 0.99 \\              
E7         &ae806038010   &  2011-06-20 & 13 08 27.43 & +27 53 06.0 & 14  & 0.99 \\               
NW1    & ae806039010  &  2011-06-20   &  12 59 14.26 & +28 10 07.0 & 7 & 0.88  \\               
NW3     & ae806040010   &  2011-06-20  & 12 58 13.70 & +28 40 10.9 & 9 & 0.85  \\     
NW3.5  & ae806040020 & 2011-12-22     & 12 58 06.94 & +28 44 19.0  & 17 & 0.85 \\          
NW4       & ae806041010  & 2011-06-20 & 12 57 43.20 & +28 54 43.2 & 19 & 0.88  \\             
NW5      &ae806042010  & 2011-06-21   & 12 57 12.89 & +29 08 43.4 & 18 & 0.95 \\
NW5.5    & ae806043010 &   2011-06-22 & 12 56 56.21 & +29 16 51.6 & 14 & 0.97   \\              
NW6        & ae806044010 & 2011-06-22  & 12 56 37.08 & +29 25 05.9 & 10  & 1.00 \\               
NW6.5    & ae806045010 &  2011-06-22 &  12 56 18.58 & +29 33 08.3 & 16 & 1.03 \\               
NW7        & ae806046010 & 2011-06-23  & 12 56 00.84 & +29 41 27.2 & 13 & 1.03   \\               
SW1        & ae806047010 & 2011-06-23  & 12 57 44.45 & +27 43 35.8 & 7 & 0.88  \\              
SW5.5    & ae806048010 & 2011-06-23  &  12 54 34.75 & +27 05 10.7 & 15 & 0.78  \\               
SW6       & ae806049010  & 2011-06-24 &  12 54 06.34 & +26 59 42.7 & 11  & 0.77 \\              
SW6.5    & ae806050010 & 2011-06-24 &   12 53 38.35 & +26 52 23.9 & 16 & 0.75   \\               
SW7     & ae806051010 &  2011-06-25  &  12 53 10.46  & +26 46 26.0 & 13 & 0.75  \\    
    \hline
NE4       & ae806021010 & 2011-12-06 &  13 03 24.84 &	+28 30 40.3  & 20 & 0.93  \\               
NE5        & ae806022010 & 2011-12-06  &  13 04 40.61 &	+28 42 33.8 & 26 & 0.95  \\              
NE6    &  ae806020010 & 2011-12-05  & 13 05 56.02 & +28 54 13.3 & 40 & 1.01  \\               
W4       & ae806025010 & 2011-12-09 &  12 55 53.76	& +28 30 38.9  & 17  & 0.85 \\              
W5    & ae806024010 & 2011-12-08  &   12 54 39.07 & +28 42 39.2  & 27 & 0.94   \\               
W6     & ae806023010 &  2011-12-07  &   12 53 22.66 &	+28 54 13.3  & 45 & 1.00  \\    
\hline
C0       & ae801097010 & 2006-05-31 &  12 59 42.41 & +27 54 22.0  & 164 & 0.91   \\               
W0        & ae801044010 & 2006-05-30  &  12 58 46.54&	+27 56 47.8 & 73 & 0.90   \\              
W1    & ae802084010 & 2007-06-21  &  12 57 22.27 & +28 08 25.1 & 29  & 0.87  \\         
NW2 & ae802082010 & 2007-06-19 & 12 58 31.32 &	+28 23 38.0 & 48 & 0.85  \\      
SW3       &  ae802047010 & 2007-12-02 &  12 57 01.39	& +27 34 17.0  & 25  & 0.84  \\              
SW4   & ae802048010 & 2007-12-04 &   12 56 06.02 & +27 25 30.7 & 30 & 0.84   \\               
SW4.5     &ae803051010  & 2008-12-23   &  12 55 27.10	& +27 17 17.9  & 176 & 0.81  \\    
S4.5    & ae805079010  &  2010-06-01  &   12 57 33.43 &	+26 55 34.0  & 85 & 0.74  \\    
\hline
  \end{tabular} 
\end{table*}

\newpage

\end{document}